\documentstyle[12pt,epsf]{article}

\begin{document}

\font\fortssbx=cmssbx10 scaled \magstep2
\hbox to \hsize{
\includegraphics{uwlogo.ps}
\hskip.5in \raise.1in\hbox{\fortssbx University of Wisconsin - Madison}
\hfill$\vcenter{\hbox{\bf MADPH-95-868}
                \hbox{October 1995}}$ }
\vskip 2cm
\begin{center}
\Large
{\bf Fermion Confinement by a Relativistic Flux Tube} \\
\vskip 0.5cm
\large
M. G. Olsson and  Sini\v{s}a Veseli \\
\vskip 0.1cm
{\small \em Department of Physics, University of Wisconsin, Madison,
	\rm WI 53706} \\
\vspace*{+0.2cm}
 Ken Williams \\
{\small \em Continuous Electron Beam Accelerator Facility \\
	Newport News, VA 29606, USA \\
	and \\
\vspace*{-0.2cm}
Physics Department, Hampton University, Hampton, VA 29668}
\end{center}
\thispagestyle{empty}
\vskip 0.7cm

\begin{abstract}
We formulate the description of the dynamic confinement of a single
fer\-mi\-on by a flux tube. The range of applicability of
this model extends from the
relativistic corrections
of a slowly moving quark to the
ultra-relativistic motion in a heavy-light meson. The reduced Salpeter
equation, also known as the no-pair equation, provides the framework
for our discussion. The Regge structure is that of a Nambu string
with one end fixed. Numerical solutions are found giving very good
fits to heavy-light meson masses. An Isgur-Wise function
with a zero recoil slope of $\xi'(1)\simeq -1.26 $ is
obtained.
\end{abstract}

\newpage
\section{Introduction}

The relativistic flux tube (RFT) model's \cite{bib:lacourse}
major success has been its ability to account for  leading
relativistic corrections in heavy quark bound states and also
to give the Regge structure of the Nambu string.
The ``Wilson loop low velocity expansion'' of Eichten and Feinberg
\cite{bib:eichten} provides a rigorous non-perturbative framework
for the discussion of these corrections.
 The spin-independent
relativistic
corrections were computed in an extension of this program
\cite{bib:prosperi}.
The spin-independent
QCD relativistic corrections are accounted for in a simple
dynamical picture by the RFT model
\cite{bib:olson}. The  spin-dependent long range corrections
have also been considered in the RFT model \cite{bib:fends}.
In the present work we show that the spin-orbit interaction
obtained as a relativistic correction, differs from the ``Thomas term''
conventionally associated with QCD \cite{bib:gromes}.

Yet to be completely explored is the ultra-relativistic sector
of the RFT model with fermions, although the spinless quark results
have been encouraging \cite{bib:lacourse,bib:collin,bib:aft}.
In fact, with reasonable assumptions the Wilson area law
 exactly reproduces the original
spinless RFT model \cite{bib:dubin,bib:gubankova}.
One of the attractive features of the RFT model
is that ultra-relativistic quark motion has little influence
on meson's rotational dynamics. For large angular momenta the
tube's momentum and energy dominate
and  the RFT reduces to the Nambu string. The RFT flux tube can
also stretch radially which involves
no change in tube's momentum. For s-waves  it then  resembles a
potential model. If both radial and angular motions are present
Coriolis force couples the radial and
rotational motions and implies
a straight leading Regge trajectory accompanied by parallel
equally spaced
daughter trajectories \cite{bib:lacourse,bib:collin}.

In recent work \cite{bib:potconf} we have examined the motion of a fermion
in a confining central potential field and found that the Dirac
equation is appropriate for scalar confinement
while the ``no-pair'' variant of the Dirac equation (or
reduced Salpeter equation),
must be used for a Lorentz vector confining potential. Since the RFT
acts like  vector confinement for low angular momenta the Salpeter
equation is relevant.

Our formalism for fermionic
quark confinement is unusual in that the confinement
is introduced into the kinetic  rather than in the usual interaction
term. The flux tube contributes to both
energy and momentum to the dynamics so it makes little sense to consider
it as a ``potential'' type interaction. By a covariant
substitution we add the tube to the quark momentum and energy.
We may equivalently view this as a ``minimal
substitution'' of a vector interaction field. The result
nicely reduces to the Nambu string in the limit in which
the quark moves ultra-relativistically.

In Section \ref{sec:ftube} we discuss the quantized form of the tube operator
and the flux tube transformation. Starting from
the reduced Salpeter (or no-pair) equation, we develop
the radial equations with tube confinement
in Section \ref{sec:radeq}. A semi-relativistic reduction
is presented in Section \ref{sec:nrred}. In Section \ref{sec:reg}
the Regge structure of the RFT model for an ultra-relativistic
fermion is determined,
yielding a straight
trajectory with slope $\alpha'= \frac{1}{\pi a}$. In Section
\ref{sec:results} we present our numerical
results. We obtain an excellent fit to the observed particle spectrum and
compute the
Isgur-Wise function for the semileptonic
$\bar{B}\rightarrow D^{(*)}l\bar{\nu}_{l}$ decays. Our conclusions are found
in Section \ref{sec:conclusions}.

\section{ Flux tube transformation}
\label{sec:ftube}

{}From the Lagrangian formulation of the spinless RFT model \cite{bib:lacourse}
it follows that the canonical linear momentum
is sum of quark and the tube parts
\begin{eqnarray}
{\bf p} &=& {\bf p}_{q} + {\bf p}_{t}\ ,
\end{eqnarray}
where
\begin{eqnarray}
p_{q} &=& \frac{L_{q}}{r} = W_{r} \gamma_{\perp} v_{\perp}\ , \\
p_{t} &=& \frac{L_{t}}{r} = 2 a r F(v_{\perp})\ ,  \label{new1}
\end{eqnarray}
with
\begin{equation}
F(v_{\perp})=\frac{1}{4v_{\perp}}(\frac{\arcsin{v_{\perp}}}{v_{\perp}}
-\frac{1}{\gamma_{\perp}}) \ ,
\end{equation}
and the tube  energy is found to be
\begin{equation}
H_{t}= ar\frac{\arcsin{v_{\perp}}}{v_{\perp}}\ .
\end{equation}
In these expressions
$\gamma_{\perp} = \frac{1}{\sqrt{1-v_{\perp}^{2}}}$ and
$W_{r}=\sqrt{p_{r}^{2}+m^{2}}$.

In an earlier work \cite{bib:fends} the flux tube with fermionic ends
was introduced through the free Dirac Hamiltonian by the covariant
transformation,
\begin{equation}
p^\mu \to p^\mu-p_t^\mu \  , \label{eq:co-trans}
\end{equation}
where
\begin{eqnarray}
p_t^\mu &=& (H_t, {\bf p}_t) \  ,\\
{\bf p}_t &=& (-\hat{{\bf r}} \times \hat{{\bf L}} ) p_t  =
- \hat{{\bf r}} \times \frac{{\bf L}_{t}}{r}\ .
\end{eqnarray}
Here, unlike in \cite{bib:fends},
we want to quantize this classical transformation
at the outset. The prescription we follow is to
first quantize $H_t$, $p_t$, and $p_{q}$
by symmetrization of the
classical expressions \cite{bib:lacourse,bib:aft}, so that
\begin{eqnarray}
H_{t}&=&\frac{a}{2} \{r,\frac{\arcsin{v_{\perp}}}{v_{\perp}}\}\ ,
\label{eq:htube}\\
p_{t}&=&  \frac{L_{t}}{r} = a \{r ,F(v_{\perp})\}\  ,\label{eq:ptube}\\
p_{q} &=& \frac{L_{q}}{r} = \frac{1}{2}\{W_{r},\gamma_{\perp}v_{\perp}\}\ ,
\end{eqnarray}
with $\{A,B\}=AB+BA$ and
$p_{r}^{2}=-\frac{1}{r}\frac{\partial^{2}}{\partial r^{2}}r$.
The quantum operators corresponding to ${\bf p}_{t}$ and ${\bf p}_{q}$  are
then
\begin{eqnarray}
{\bf p}_{t} &=&
\frac{1}{2}(\frac{{\bf L}_{t}}{r} \times \hat{{\bf r}}
-\hat{{\bf r}} \times \frac{{\bf L}_{t}}{r}) \ \\
{\bf p}_{q} &=& \hat{{\bf r}} p_{r} +
\frac{1}{2}(\frac{{\bf L}_{q}}{r} \times \hat{{\bf r}}
-\hat{{\bf r}} \times \frac{{\bf L}_{q}}{r})\ \label{eq:ptot}
\end{eqnarray}
where $p_{r}=-\frac{i}{r}\frac{\partial}{\partial r}r$ is hermitian.
{}From  ${\bf p}={\bf p}_{q} + {\bf p}_{t}$  and the identity
\begin{equation}
\hat{{\bf r}} p_{r} + \frac{1}{2}(\frac{{\bf L}}{r} \times \hat{{\bf r}}
-\hat{{\bf r}} \times \frac{{\bf L}}{r}) =
\hat{{\bf r}} (-i\frac{\partial}{\partial r}) -
\hat{{\bf r}} \times \frac{{\bf L}}{r}\ ,
\end{equation}
we obtain the usual
\begin{eqnarray}
{\bf p} &=& \hat{{\bf r}} (-i\frac{\partial}{\partial r}) - \hat{{\bf r}}
\times \frac{{\bf L}}{r}\ .
\label{eq:pfict}
\end{eqnarray}
The full flux tube transformation (\ref{eq:co-trans})
of the free Dirac Hamiltonian is then
given by
\begin{equation}
H_{0}({\bf p}) \rightarrow H_{0}({\bf p}) + H_{t} -\mbox{\boldmath$\alpha$}
\cdot {\bf p}_{t}\ ,
\label{eq:ftr}
\end{equation}
where
\begin{equation}
H_{0}({\bf p}) = \mbox{\boldmath $\alpha$}\cdot {\bf p} +\beta m\ .
\end{equation}
We choose to work with the reduced Salpeter
equation \cite{bib:salpeter,bib:gara}
in the limit where the heavy antiquark mass is infinite
\cite{bib:long}.
In coordinate space it is given by
\begin{equation}
\Lambda_{+}[E_{q}-H_{0}({\bf p}) -V(r)]\Lambda_{+}\Psi({\bf r}) = 0\  ,
\label{eq:rsalp}
\end{equation}
and also known as the no-pair equation \cite{bib:hardekopf}.
Here, $E_{q}=M-m_{\bar{Q}}$ is the energy of the light degrees
of freedom, $V(r)$ is the short range vector interaction, while
$\Lambda_{+}$ is the positive energy projection operator
\begin{equation}
\Lambda_{+}=\frac{E_{0} + H_{0}}{2E_{0}}\  ,
\end{equation}
with
\begin{equation}
E_{0}=\sqrt{{\bf p}^{2}+m^{2}}\ .
\end{equation}
The flux-tube transformation (\ref{eq:ftr})
on equation (\ref{eq:rsalp})
 results in the no-pair equation describing heavy-light mesons
with the flux tube accounting for the long range interaction
\begin{equation}
[E_{q}-H_{0}({\bf p}) - \Lambda_+
( V(r)+ H_{t} -\mbox{\boldmath$\alpha$} \cdot {\bf p}_{t}   ) \Lambda_{+}]\Psi
({\bf r}) = 0 \  .
\label{eq:np}
\end{equation}

\section
{Radial equations}
\label{sec:radeq}

Coupled radial eigenvalue equations follow from the spherical symmetry of
equation (\ref{eq:np}). As in \cite{bib:potconf} the solutions are of the
form
\begin{equation}
\Psi^k_{jm}({\bf r}) = \left(
\begin{array}{c}
f^k_j(r) {\cal Y }^k_{jm}(\hat{{\bf r}}) \\
i g^k_j(r) {\cal Y }^{-k}_{jm}(\hat{{\bf r}})
\end{array}
\right)
 \equiv {\cal Y }\hspace{-2.35mm}{\cal Y } \left(
\begin{array}{c}
f^k_j   \\
 g^k_j
\end{array}
\right)\  ,\label{eq:wf}
\end{equation}
where
\begin{equation}
{\cal Y }\hspace{-2.35mm}{\cal Y } = \left(
\begin{array}{cc}
{\cal Y }^k_{jm} & 0 \\
0 &  i {\cal Y }^{-k}_{jm}
\end{array}
\right)  \  .
\end{equation}
The Dirac quantum number $k$ labels the bound state
and is defined by \cite{bib:gross}
\begin{equation}
k=\pm(j+\frac{1}{2})\ .\label{eq:defk}
\end{equation}
The identities
\begin{eqnarray}
\mbox{\boldmath$\sigma$} \cdot {\bf L}  {\cal Y }^k_{jm} &=& -(k+1) {\cal Y
}^k_{jm}\  , \\
\mbox{\boldmath$\sigma$} \cdot \hat{{\bf r}} {\cal Y }^k_{jm} &=& - {\cal Y
}^{-k}_{jm} \  ,
\end{eqnarray}
lead to
\begin{eqnarray}
\mbox{\boldmath$\sigma$} \cdot {\bf p}  {\cal Y }^{\pm k}_{jm} (\hat{{\bf r}})
&=&
i {\cal Y }^{\mp k}_{jm} (\hat{{\bf r}})  D_{\pm}  \  ,\\
\mbox{\boldmath$\sigma$} \cdot {\bf p}_t  {\cal Y }^{\pm k}_{jm} (\hat{{\bf
r}}) &=& \mp
i  {\cal Y }^{\mp k}_{jm} (\hat{{\bf r}})  T_t \ ,
\end{eqnarray}
where
\begin{eqnarray}
D_{\pm} &=& \pm \frac{k}{r} + (\frac{\partial }{\partial r} +\frac{1}{r} )
\ ,\label{eq:d}\\
T_{t} &=& \frac{1}{2} \left[ \frac{(1-k)}{\sqrt{-k(1-k)}} p^{-k}_t
-\frac{(1+k)}{\sqrt{k(1+k)}} p^k_t \right] \ . \label{eq:tt}
\end{eqnarray}
The no-pair equation can now be expressed in radial form. Starting from
(\ref{eq:np}) and the wave function (\ref{eq:wf}), we use the above
relations  to obtain matrix radial no-pair equations
\begin{equation}
(E_{q} - h\hspace{-1.85mm}h_{0}  - I\hspace{-1.65mm}L I\hspace{-1.65mm}I
I\hspace{-1.65mm}L )\left(
\begin{array}{c}
f^k_j   \\
 g^k_j
\end{array}
\right) = 0\ ,\label{eq:rnpeq}
\end{equation}
with  definitions
\begin{eqnarray}
h\hspace{-1.85mm}h_{0} &\equiv & \left(
\begin{array}{cc}
m & -D_- \\
D_+ & -m
\end{array}
\right) \ ,\label{eq:h0}\\
I\hspace{-1.65mm}L &\equiv & \left(
\begin{array}{cc}
\lambda_+ & -\frac{1}{2E_{0}^{+}} D_- \\
D_+ \frac{1}{2E^{+}_{0}} & \lambda_-
\end{array}
\right) \  ,\label{eq:l}\\
I\hspace{-1.65mm}I &\equiv & \left(
\begin{array}{cc}
V(r)+H_{t}^{k} & T_{t} \\
T_{t} & V(r)+H_{t}^{-k}
\end{array}
\right)\  .\label{eq:i}
\end{eqnarray}
In the above expressions
\begin{eqnarray}
\lambda_{\pm} &=& \frac{E^{\pm }_0 \pm m }{2 E^{\pm}_0 } \ ,
\label{eq:lambda}\\
E^{\pm}_{0} &=& \sqrt{m^{2} - D_{\mp}D_{\pm}}\ . \label{eq:e0}
\end{eqnarray}
{}From (\ref{eq:i}) we observe  that the tube energy enters much
like the time component of the short range
Lorentz vector potential $V(r)$, which we
take to be
\begin{equation}
V(r)=-\frac{\kappa}{r}\ .
\end{equation}
Finally, the only unknown operators in
(\ref{eq:rnpeq}) are $v_{\perp}^{\pm k}$,
which are contained in $p_{t}^{\pm k}$
 and $H_{t}^{\pm k}$.
These will be determined from the heavy-light orbital angular momentum
equation as in the spinless RFT model \cite{bib:aft}. From
\begin{equation}
\left[ \frac{L}{r}
 = \frac{1}{2}\{W_{r},\gamma_{\perp}v_{\perp}\}+
a \{r,F(v_{\perp})\}\right] \Psi({\bf r})
\ ,\label{eq:angmom}
\end{equation}
and using $L{\cal Y }_{jm}^{k}(\hat{{\bf r}})=
\sqrt{k(k+1)}{\cal Y }_{jm}^{k}(\hat{{\bf r}})$,
one immediately finds equations that determine the  $v_{\perp}^{\pm k}$
operators:
\begin{eqnarray}
&\left[  \frac{\sqrt{k(k+1)}}{r}  =
\frac{1}{2}\{W_{r},\gamma_{\perp}^{k}v_{\perp}^{k}\}+
a \{r,F(v_{\perp}^{k})\}\ \right]
f_{j}^{k}(r){\cal Y }_{jm}^{k}(\hat{{\bf r}}) \  ,& \label{eq:angmom1} \\
&\left[ \frac{\sqrt{-k(-k+1)}}{r}  =
\frac{1}{2}\{W_{r},\gamma_{\perp}^{-k}v_{\perp}^{-k}\}+
a \{r,F(v_{\perp}^{-k})\}\ \right]
g_{j}^{k}(r){\cal Y }_{jm}^{-k}(\hat{{\bf r}})\ .& \label{eq:angmom2}
\end{eqnarray}
The numerical technique used to solve for $v_{\perp}$ is discussed in detail
elsewhere \cite{bib:lacourse,bib:aft}.

\section{Semi-relativistic reduction}
\label{sec:nrred}

For comparison with previous results we decouple (\ref{eq:rnpeq})
 and make the non-re\-la\-ti\-vistic reduction to order $(v/c)^4 $
(ignoring the short-range V(r) component),
\begin{eqnarray}
H = E_q &=& m + \frac{1}{2 E^{+}_{0}}
 \Bigl[(E^+_{0} +m) H^k_t (E^+_{0} +m)+(E^+_{0} +m)  T_t D_+  \nonumber \\
&-& D_- T_t  (E^+_{0}+m)
 - D_- H^{-k}_t D_+ \Bigr] \frac{1}{2 E^+_{0}}
- \frac{D_- D_+}{E^+_{0}+m}  \nonumber \\
&+& \frac{1}{2 E^+_{0}} \Bigl[-(E^+_{0} +m)  H^k_t D_- D_+ +(E^+_{0} +m) T_t
(E^-_{0} -m) D_+ \nonumber \\
&+& D_- T_t D_- D_+ -D_- H^{-k}_t (E^-_{0} -m) D_+ \Bigr] \frac{1}{2 E^+_{0}
(E^+_{0} +m)} \ .\label{eq:nred}
 \end{eqnarray}
{}From (\ref{eq:htube}) and (\ref{eq:ptube})  we find
(with $L\simeq mv_{\perp}r$)
\begin{eqnarray}
H_t &\simeq & a r + \frac{1}{6} a r v_{\perp } \simeq a r +
\frac{a L^2}{6 m^2 r} \  ,\\
p_t &\simeq & \frac{1}{3} a r v_{\perp } \simeq \frac{a L}{3 m} \ ,
\label{eq:ltr}
\end{eqnarray}
and from (\ref{eq:nred}) it then follows
\begin{eqnarray}
H &\simeq & m-\frac{D_-D_+}{2m} -\frac{D_-D_+D_-D_+}{8m^3} +
H^k_t \nonumber \\
& +&
\frac{1}{4m^2} \Bigl( D_-D_+H^k_t-D_- H^{-k}_t D_+ \Bigr) -\frac{1}{2m} \Bigl(
D_- T_t -T_t D_+ \Bigr) \ .
\end{eqnarray}
Taking expectation values, and  using
\begin{eqnarray}
k(k+1) &\rightarrow & L^2  \ , \\
-(k+1) &\rightarrow & 2 {\bf S} \cdot {\bf L} \ ,\\
D_-D_+ &\rightarrow & -  p^2 \ ,
\end{eqnarray}
we obtain the effective Hamiltonian
\begin{equation}
H =  m+\frac{p^2}{2m} -\frac{p^4}{8 m^3} +ar -\frac{a}{6 m^2}
\frac{{\bf S} \cdot {\bf L} }{r} -\frac{a}{6 m^2 }\frac{L^2}{r} -
\frac{a}{12 m^2r}  \ .\label{eq:red}
\end{equation}
Physical interpretation can be established
for each of the terms of the above effective Hamiltonian.
The first three terms
result from the expansion
of the relativistic energy $\sqrt{p^{2}+m^{2}}$, and the fourth
is the static tube energy.
The spin-orbit terms is the net result of the kinematic Thomas precession
and tube energy and momentum contributions. As already mentioned, our
result differs from the one conventionally found from QCD \cite{bib:gromes}.

The $L^{2}$ term in (\ref{eq:red}) has a nice
interpretation as pointed out previously \cite{bib:olson}. The tube's
explicit contribution to the angular momentum is the sixth term of
(\ref{eq:red}). There is also a hidden contribution from the
$p^{2}/2m$
term. The rotational part of ${\bf p}$ yields to
leading tube order ($p_{t}\ll p_{q}$)
\begin{eqnarray}
\frac{p^{2}}{2m} &\simeq& \frac{p_{q}^{2}}{2m} +
\frac{p_{q}}{m}\frac{L_{t}}{r}\nonumber \\
&=& \frac{p_{q}^{2}}{2m} + (\frac{L}{mr})(\frac{aL}{3m})\ ,
\end{eqnarray}
where we have used $L\simeq p_{q}r$ and $L_{t}/r\simeq aL/3m$
from (\ref{eq:ltr}). The total tube contribution
to the $L^{2}$ part of the Hamiltonian is then
\begin{equation}
\Delta H_{t} = \frac{aL^{2}}{m^{2}r} (\frac{1}{3}-\frac{1}{6})
= \frac{aL^{2}}{6m^{2}r}\ .
\end{equation}
The above is exactly the rotational energy of a
rod of a mass $ar$ and length $r$ rotating
with angular velocity $\omega=L/mr^{2}$,
\begin{eqnarray}
E_{t}^{rot} &=& \frac{1}{2} I\omega^{2}\nonumber \\
&=&\frac{1}{2}(\frac{1}{3}ar) r^{2} \frac{L^{2}}{m^{2}r^{4}}=
\frac{aL^{2}}{6m^{2}r}\ .
\end{eqnarray}
The final term in (\ref{eq:red}) is
a Darwin like term and it is sensitive to how the Hamiltonian
is symmetrized. If one symmetrizes at the end as in previous
work \cite{bib:prosperi,bib:fends}, a different result is obtained.

\section{Regge  structure}
\label{sec:reg}

A confined particle in a sufficiently large angular momentum state moves
ultra relativistically. For linear confinement
we expect linear Regge trajectories from
classical considerations \cite{bib:aft,bib:goebel}. In the flux tube
model the angular momentum and
rotational energy of the quark
are negligible compared to those of the tube. The slope of the
Regge trajectory can be found from the model without a detailed solution
of the wave equation. To determine the Regge slope
directly from the no-pair tube equation
we note from (\ref{new1}) and (\ref{eq:tt}) that ($m\rightarrow 0,\
p_{r}\rightarrow 0,\ v_{\perp} \rightarrow 1$)
\begin{eqnarray}
p_{t}^{ k} &\stackrel{|k|\gg 1}{\longrightarrow} & \frac{|k|}{r}  \ ,\\
T_{t}&\stackrel{|k|\gg 1}{\longrightarrow}& - \hat{k} p_{t}\ ,
\end{eqnarray}
with $ \hat{k}  = \frac{k}{|k|} $ being the sign of $ k $,
while from (\ref{eq:htube}) it follows
\begin{equation}
H_{t}^{k} \stackrel{|k|\gg 1}{\longrightarrow} \frac{ar\pi}{2}\ .
\end{equation}
We also have
\begin{eqnarray}
E_{0}^{\pm}& \stackrel{|k|\gg 1}{\longrightarrow} & p_{t} \ ,\\
\lambda^{\pm} &\stackrel{|k|\gg 1}{\longrightarrow} & \frac{1}{2}\ ,
\end{eqnarray}
 so that in the limit $|k|\gg 1$ we obtain
\begin{eqnarray}
h\hspace{-1.85mm}h_{0} &\stackrel{|k|\gg 1}{\longrightarrow}&
p_{t}\left( \begin{array}{cc}
0 & \hat{k}  \\
\hat{k} & 0\end{array}\right)\  ,\\
I\hspace{-1.65mm}L &\stackrel{|k|\gg 1}{\longrightarrow}&
\frac{1}{2}\left( \begin{array}{cc}
1 & \hat{k} \\
\hat{k} & 1\end{array}\right)\  ,\\
I\hspace{-1.65mm}I &\stackrel{|k|\gg 1}{\longrightarrow}&
\left( \begin{array}{cc}
H_{t} &-\hat{k}p_{t} \\
-\hat{k}p_{t} & H_{t}\end{array}\right)\ .
\end{eqnarray}
Multiplying out $I\hspace{-1.65mm}LI\hspace{-1.65mm}II\hspace{-1.65mm}L $ then
yields
\begin{equation}
E_{q}=I\hspace{-1.65mm}LI\hspace{-1.65mm}II\hspace{-1.65mm}L +
h\hspace{-1.85mm}h_{0} \simeq \frac{1}{2}
\left( \begin{array}{cc}
H_{t}-p_{t} & (H_{t}+p_{t})\hat{k} \\
(H_{t}+p_{t})\hat{k} & H_{t}-p_{t} \end{array}\right)\ .\label{eq:lil}
\end{equation}
Independent of the sign of $\hat{k}$, the eigenvalues of (\ref{eq:lil})
are
\begin{eqnarray}
E_{q+} &=& H_{t} \simeq \frac{ar\pi}{2}\  ,\label{eq:eplus}\\
E_{q-} &=& -p_{t}\simeq -\frac{|k|}{r}\ .
\end{eqnarray}
The positive energy solution has the proper Regge
behavior. From (\ref{eq:angmom}) we find the total orbital
angular momentum of the system  in the ultra-relativistic limit to be
\begin{equation}
L\stackrel{|k|\gg 1}{\longrightarrow} \frac{a\pi r^{2}}{4}\simeq |k|\ .
\end{equation}
Eliminating $r$ and using (\ref{eq:eplus})
we obtain the ratio of angular momentum
to the square of the
energy of the light degrees of freedom,
\begin{equation}
\frac{|k|}{E_{q}^{2}}\stackrel{|k|\gg 1}{\longrightarrow}\frac{1}{\pi a}\ .
\end{equation}
This is the same slope that one would obtain from a Nambu
string with one end fixed.

\section{Numerical results}
\label{sec:results}

By expanding $f_{j}^{k}$ and $g_{j}^{k}$ in terms of a complete
set of basis states, and then truncating the expansion to the
first $N$ states we can transform the angular momentum
equations (\ref{eq:angmom1}-\ref{eq:angmom2}) into two $N\times N$
matrix equations from which matrices for $v_{\perp}^{\pm k}$
operators are found \cite{bib:aft}. Also, the radial no-pair equation
becomes
 a $2N\times 2N$ matrix equation, which we solve using the
Galerkin
variational method. The quasi-Coulombic basis states (which depend on
a scale
parameter $\beta$), and all
matrix elements used are described in \cite{bib:potconf}. In Fig.
\ref{fig:bplt} we show the dependence on  $\beta$ for
 the three lowest positive
and three highest negative energy states, with the particular choice
$m=0.3\ GeV,\ k=-1,\ j=\frac{1}{2}, \ a=0.2\ GeV^{2}$, and
$\kappa = 0.5$. As we increase the number of states $N$, the plateau
where the positive eigenvalues are stable enlarges.

The Regge slope of the no-pair equation with the flux tube
was shown in Section \ref{sec:reg}
 to be $\alpha'=\frac{1}{\pi a}$. Our
numerical solution agrees as shown
in Fig. \ref{fig:regge}. In this figure we illustrate
the leading trajectories for
the two light degrees of freedom parity
states
corresponding to $k=\pm(j+\frac{1}{2})$ for  $m_{u,d}=0.3\ GeV,
\ a=0.2\ GeV^{2}$ and $\kappa = 0.5$. We also
show several
daughter trajectories
corresponding to radial excitations.

In order to recover the universal Regge slope,
 we fix $a$
to be $0.2\ GeV^{2}$, and we also choose $m_{u,d}=0.3\ GeV$
and fit to the spin averaged
heavy-light meson states. Our result is shown in Table \ref{tab:hl}, and
parameters of the fit are
\begin{eqnarray}
m_{u,d}&=&0.300 \ GeV \ {\rm (fixed)}\ ,\nonumber \\
m_{s}&=&0.581\ GeV\ ,\nonumber \\
m_{c}&=&1.286\ GeV\ ,\nonumber \\
m_{b}&=&4.621\ GeV\ ,\label{eq:fit}\\
a&=&0.200 \ GeV^{2}\ {\rm (fixed)}\ ,\nonumber \\
\kappa&=&0.641\ . \nonumber
\end{eqnarray}
The agreement of the
fitted levels to experiment
is excellent, all spin-averaged states are reproduced with
errors of about  $5\ MeV$.

Finally we use the wavefunctions
with parameters of  (\ref{eq:fit}) to evaluate the Isgur-Wise function
for the semileptonic $\bar{B}\rightarrow D^{(*)}l\bar{\nu}_{l}$ decays
using \cite{bib:zalewski}
\begin{equation}
\xi(\omega)
=
\frac{2}{\omega + 1}
\langle\
j_{0}(2E_{q}\sqrt{\frac{\omega-1}{\omega+1}}r)
\rangle\ , \label{eq:iwf}
\end{equation}
where
\begin{equation}
\langle A \rangle =
\int_{0}^{\infty}dr\  r^2 R(r)A(r)R(r)\ .
\end{equation}
 As shown on
the Fig. \ref{fig:iwf}, the agreement with
ARGUS \cite{bib:arg} and CLEO II \cite{bib:cleo} data is excellent.
For the seven CLEO II data points, this IW function has
 $\chi^{2}$ per degree of freedom
of 0.4 (corresponding to about $90\% \ CL$).
To calculate the slope, we use \cite{bib:iwf} expression
\begin{equation}
\xi'(1) = -(\frac{1}{2}+\frac{1}{3}E_{q}^{2}
\langle r^{2}\rangle) \ .\label{eq:first}\\
\end{equation}
For the light quark mass of 300 $MeV$, and using values for $a$ and
$\kappa$  from (\ref{eq:fit})
we find
\begin{equation}
\xi'(1)=-1.26\ .\label{eq:slope}
\end{equation}

We observe that slopes calculated from the different versions
of the no-pair equation (with vector confinement \cite{bib:potconf}
 or with the flux tube) are more negative
($\xi'(1)$ ranging from about -1.2 to -1.3) than the ones calculated
from the spinless RFT model or the Dirac equation with scalar confinement
($\xi'(1)\simeq -0.9$).

\section{Conclusions}
\label{sec:conclusions}

In this paper we further explore an idea \cite{bib:fends}
for describing the relativistic quantized system of a flux tube with fermionic
quarks at its ends. We work out in detail the simplest
case where one fermion has infinite mass. Our technique
for incorporating the tube into the reduced Salpeter equation
is the covariant flux tube transformation (\ref{eq:ftr}).
In
order to achieve a satisfactory description of a heavy-light system,
there are two physical requirements that have to be satisfied:
\begin{enumerate}
\item
In the ultrarelativistic limit the tube dynamics must dominate
giving a Nambu string limit.
\item
The semi-relativistic corrections must agree with rigorous QCD expectations.
\end{enumerate}
This model meets the above requirements with the possible exception
of the spin-orbit interaction.

Let us briefly review the evidence concerning the QCD
spin-orbit interaction. The conventional wisdom is that the
spin-orbit effective Hamiltonian is equivalent to pure Thomas
precession implying a coefficient $-\frac{1}{2}$ instead
of $-\frac{1}{6}$ in the ${\bf S} \cdot {\bf L}$ term of
(\ref{eq:red}). The analysis of Eichten and Feinberg \cite{bib:eichten}
does not say what the spin-orbit coefficient is although
later work by Gromes \cite{bib:gromes}  and the
Milan group \cite{bib:prosperi}, with further assumptions,
find the Thomas result. We believe this remains an open
question.

The model also yields the expected
Regge structure. For large angular momentum a Regge slope of
$\alpha' = \frac{1}{\pi a}$, characteristic for a Nambu
string with one fixed end, is obtained. A sequence of parallel equally
spaced daughter trajectories follows from radial excitations of the
quark.  By varying
the tension $a$, the Coulomb coefficient
$\kappa$ and  the quark masses we can account for the known
spin-averaged heavy-light levels. In Table \ref{tab:hl}
we show
a fit with the light quark mass and the tension
fixed at reasonable values.
We also used the s-wave wave function to compute
the Isgur-Wise function. The result agrees well with the
known experimental data, as shown in Fig. \ref{fig:iwf}.

The numerical techniques developed for the flux tube model with spinless quarks
\cite{bib:lacourse, bib:aft} can be directly applied in the fermionic
case since quantization of the orbital angular momentum sector
is independent of the spin part. The extension to arbitrary
two fermion
systems seems to be free of any intrinsic difficulties.

\vskip 1cm
\begin{center}
ACKNOWLEDGMENTS
\end{center}
This work was supported in part by the U.S. Department of Energy
under Contract Nos.  DE-FG02-95ER40896 and DE-AC05-84ER40150,
the National Science
Foundation under Grant No. HRD9154080,
and in part by the University
of Wisconsin Research Committee with funds granted by the Wisconsin Alumni
Research Foundation.

\begin{table}
\normalsize
\begin{center}
TABLES
\end{center}
\caption{ Heavy-light spin averaged states.
Theoretical
results are obtained from the no-pair equation with
the flux tube. Spin-averaged
masses are calculated in the usual way, by taking $\frac{3}{4}$
($\frac{5}{8}$) of
the triplet and $\frac{1}{4}$ ($\frac{3}{8}$) of the singlet mass for the
s(p)-waves). }
\vskip 0.2cm
\begin{tabular}{|lccccccc|}
\hline
\hline
         state
       & \multicolumn{2}{c}{spectroscopic label\hspace{+2mm}}
       & spin-averaged
       & \multicolumn{2}{c}{q. n.}
       & theory
       & error
\\
       & $J^{P}$
       & $^{2S+1}L_{J}$
       & mass (MeV)
       & $j$
       & $k$
       & (MeV)
       & (MeV)
\\
\hline
         \underline{$c\bar{u},\ c\bar{d}$ quarks}
       &
       &
       &
       &
       &
       &
       &
\\
         $\begin{array}{ll}
              		D     &  (1867) \\
   			D^{*} &  (2009) \end{array}$
       & $\begin{array}{l}
       			0^{-} \\
      			1^{-} \end{array}$
       &
	 $\left. \begin{array}{l}
			\hspace{+1.1mm}   ^{1}S_{0} \\
			\hspace{+1.1mm}   ^{3}S_{1} \end{array}\right] $
       & $1S\ (1974)$
       & $\frac{1}{2}$
       & $-1$
       & $1980$
       & $6$
\\
	 $\begin{array}{ll}
			D_{1}     & (2425) \\
			D_{2}^{*} & (2459) \end{array}$
       & $\begin{array}{l}
			1^{+} \\
			2^{+} \end{array}$
       & $\left.\begin{array}{l}
			\hspace{+0.5mm}^{1}P_{1} \\
			\hspace{+0.5mm}^{3}P_{2} \end{array}\right] $
       & $1P\ (2446)$
       & $\frac{3}{2}$
       & $-2$
       & $2440$
       & $-6$
\\
         \underline{$c\bar{s}$ quarks}
       &
       &
       &
       &
       &
       &
       &
\\
	 $\begin{array}{ll}
   			D_{s} & (1969) \\
   			D^{*}_{s}&  (2112) \end{array}$
       & $\begin{array}{l}
      			0^{-} \\
      			1^{-} \end{array}$
       & $\left. \begin{array}{l}
			\hspace{+1.1mm}    ^{1}S_{0} \\
			\hspace{+1.1mm}    ^{3}S_{1} \end{array}\right] $
       & $1S\ (2076)$
       & $\frac{1}{2}$
       & $-1$
       & $2072$
       & $-4$
\\
	 $\begin{array}{ll}
			D_{s1} & (2535) \\
		        D_{s2} & (2573) \end{array}$
       & $\begin{array}{l}
			1^{+} \\
			2^{+} \end{array}$
       & $\left.\begin{array}{l}
			\hspace{+0.5mm}^{1}P_{1} \\
		  	\hspace{+0.5mm}^{3}P_{2} \end{array}\right] $
       & $1P\ (2559)$
       & $\frac{3}{2}$
       & $-2$
       & $2563$
       & $4$
\\
         \underline{$b\bar{u},\ b\bar{d}$ quarks}
       &
       &
       &
       &
       &
       &
       &
\\
         $\begin{array}{ll}
   			B     & (5279) \\
   			B^{*} &  (5325) \end{array}$
       & $\begin{array}{l}
     	 		0^{-} \\
      			1^{-} \end{array}$
       & $\left. \begin{array}{l}
			\hspace{+1.1mm}   ^{1}S_{0} \\
 			\hspace{+1.1mm}   ^{3}S_{1} \end{array}\right] $
       & $1S\ (5314)$
       & $\frac{1}{2}$
       & $-1$
       & $5315$
       & $1$
\\
         \underline{$b\bar{s}$ quarks}
       &
       &
       &
       &
       &
       &
       &
\\
         $\begin{array}{ll}
			B_{s}     & (5374) \\
			B_{s}^{*} & (5421) \end{array}$
       & $\begin{array}{l}
			0^{-} \\
			1^{-} \end{array} $
       & $\left.\begin{array}{l}
			\hspace{+1.0mm} ^{1}S_{0} \\
			\hspace{+1.0mm} ^{3}S_{1} \end{array}\right] $
       & $1S\ (5409)$
       & $\frac{1}{2}$
       & $-1$
       & $5408$
       & $-1$
\\
\hline
\hline
\end{tabular}
\\
\label{tab:hl}
\end{table}

\begin{figure}[p]
\begin{center}
FIGURES
\vskip 2mm
\end{center}
\caption{Dependence on the basis state scale
$\beta$ of the three lowest positive energy states
and the three highest negative energy states for
the no-pair equation with flux tube. Here we have chosen
$m_{q}=0.3\ GeV,\ a=0.2\ GeV^{2},
\ \kappa=0.5,\ k=-1$ and $j=\frac{1}{2}$. The full lines correspond to
$N=25$, and dashed lines correspond to
$N=15$
basis states used.}
\label{fig:bplt}
\end{figure}

\begin{figure}
\caption{Regge
trajectories for the no-pair equation with flux tube.
Again, we have chosen $m_{u,d}=0.3\ GeV,\ a=0.2 GeV^{2}$, and
$\kappa = 0.5$. The full lines correspond to $k=-(j+\frac{1}{2})$,
and dashed lines to $k=j+\frac{1}{2}$. To ensure
that all calculated energies are correct, we used $N=50$ basis
states, and kept first 15 states.}
\label{fig:regge}
\end{figure}

\begin{figure}
\caption{The Isgur-Wise
 function for $\bar{B}\rightarrow D^{(*)}l\bar{\nu}_{l}$ decays calculated from
the
no-pair equation with tube confinement. Values for the
light quark mass $m_{u,d}$,  tension $a$ and
short range potential constant $\kappa$ are taken from
 the fit (\protect\ref{eq:fit}). For the
sake of clarity, error bars are shown only for the
CLEO II data.}
\label{fig:iwf}
\end{figure}

\begin{figure}
\vskip 7cm
\end{figure}

\clearpage

\begin{figure}[p]
\epsfxsize = 5.4in \epsfbox{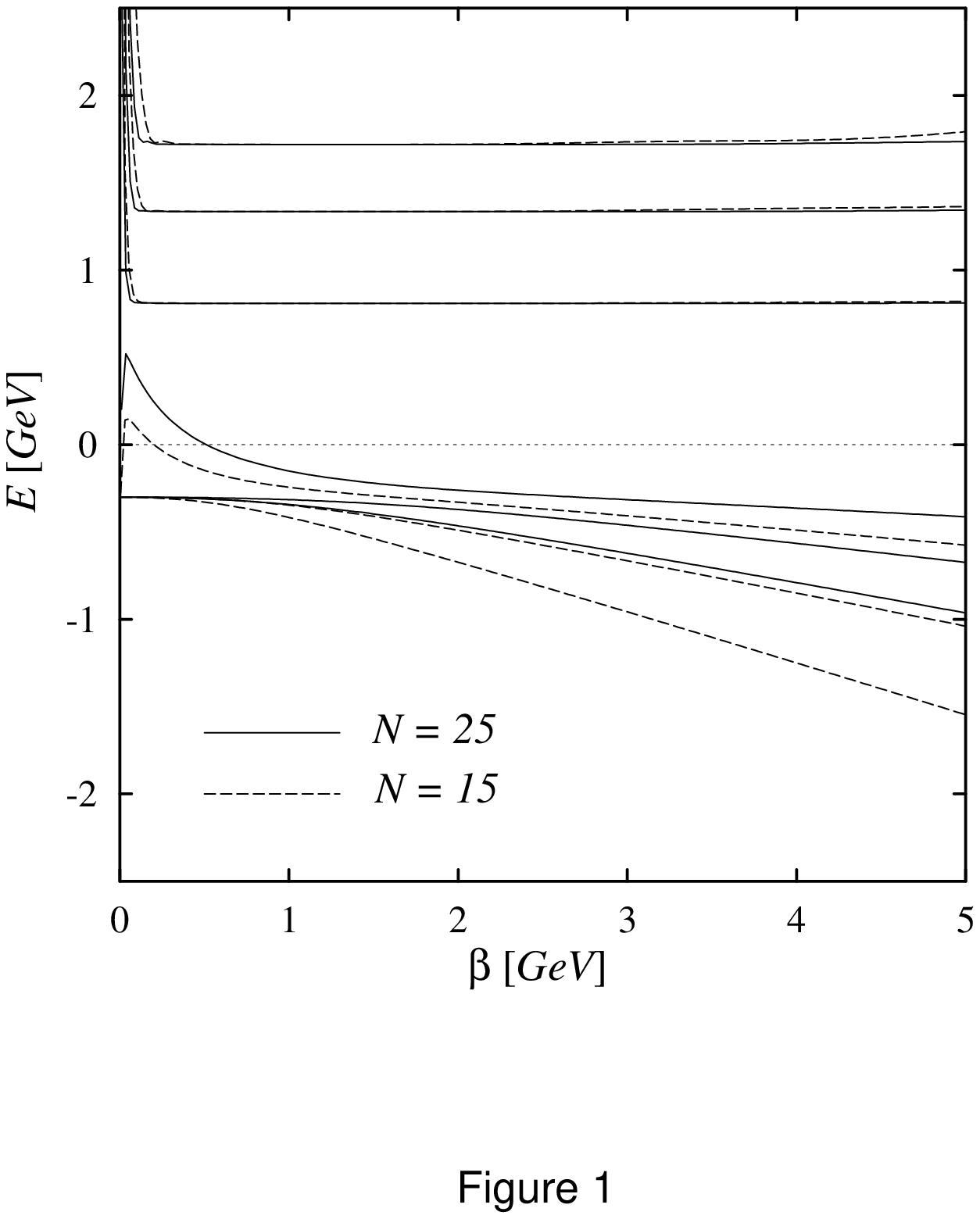}
\end{figure}

\begin{figure}[p]
\epsfxsize = 5.4in \epsfbox{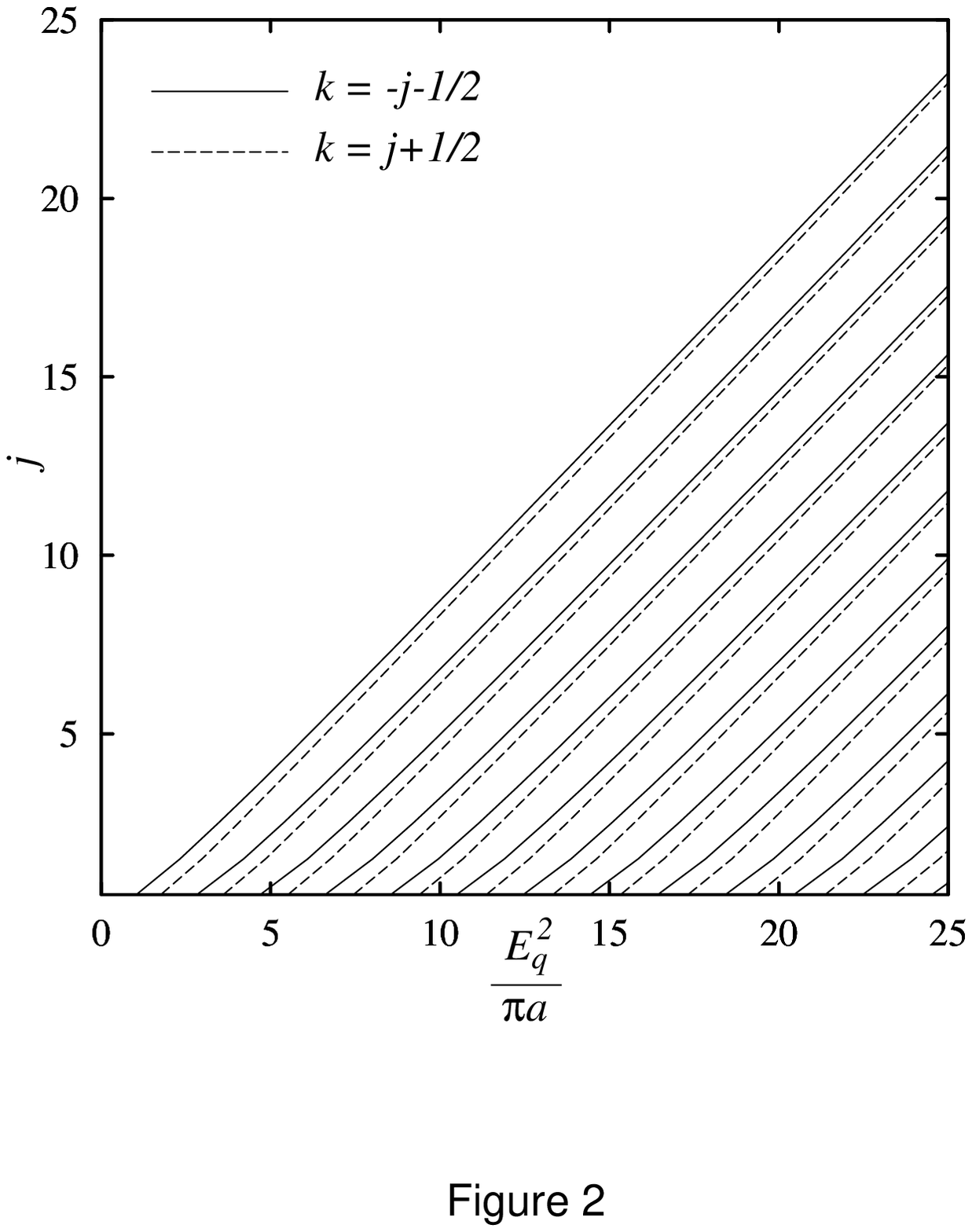}
\end{figure}

\begin{figure}[p]
\epsfxsize = 5.4in \epsfbox{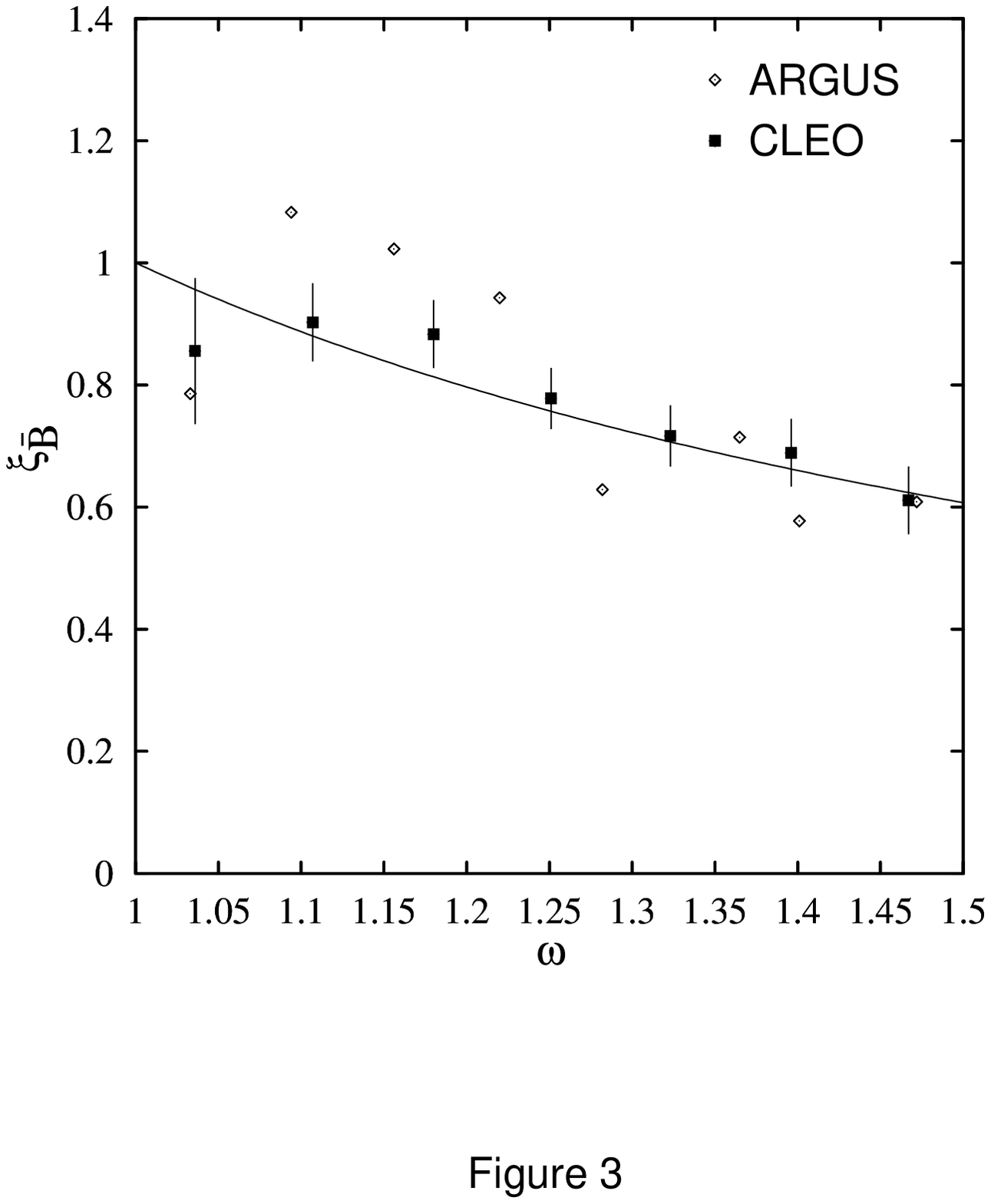}
\end{figure}

\end{document}